\documentclass[12pt]{article}
\usepackage[english]{babel}
\usepackage{epsfig}
\begin{document}

\par
{\bf Problem of oscillations presence at $CP$ violation in the
system of $K^o$ mesons}
\\

\par

\begin{center}
Kh. M. Beshtoev
\vspace{0.1cm}
\par
\vspace{0.3cm} Joint Institute for Nuclear Research, Joliot Curie
6, 141980 Dubna, Moscow region, Russia.
\end{center}

\par
Abstract

\par
In this work there are considered two approaches to  the
description of $K^o, \bar K^o$ meson transitions into $K_S
(K^o_1)$ mesons at $CP$ violation in weak interactions. The first
approach uses the  standard theory of oscillations and the second
approach supposes that ($K_S, K_L$) states which arise at $CP$
violation are normalized but not orthogonal state functions then
there arise interferences between these states but not
oscillations. It is necessary to remark that the available
experimental data are in good agreement with the second approach.
So we came to the conclusion that oscillations do not arise at
$CP$ violation in weak interactions in the system of $K^o$ mesons.
Only interference between two - $K_S, K_L$ states takes place
there.

\section{Introduction}

\par
Oscillations of $K^o$ mesons (i.e., $K^{o} \leftrightarrow \bar
K^{o}$) were theoretically \cite{1} and experimentally \cite{2}
investigated in the 50-s and 60-s. Recently there has been
achieved an understanding that these processes go as a
double-stadium process \cite{3,4,5,6}. A detailed study of $Ê^{î}$
meson mixing and oscillations is very important since the theory
of neutrino oscillations is built by analogy with the theory of
$Ê^{î}$ meson oscillations.
\par
Previously it was supposed that $P$ parity is a well number,
however, after theoretical \cite{7} and experimental \cite{8}
works it has become clear that in weak interactions $P$ parity is
violated. Then in work \cite{9} there was an advanced supposition
that in weak interactions $CP$ parity is conserved but not $P$
parity. Work \cite{10} has reported that in $K_L$ decays with a
probability of about $0.2\%$ there is two $\pi$ decay mode that is
a detection of $CP$ parity violation.

\par
A phenomenological analysis of $K^{o}$ meson processes was done in
work \cite{11} (see also \cite{12}). There non unitary
transformation and non orthogonal states where used at obtaining
$K_S, K_L$ states. It was supposed that these states arise at $CP$
violation. In work \cite{13} there was considered the same process
in the framework of the standard scheme (theory) of $K^{o}$ meson
oscillations.
\par
This work is continuation of pervious work \cite{13}. In this work
we will consider elements of the theory of $K^o$ meson
oscillations at strangeness ($S$) and $CP$ violations then the
case of $CP$ violation at absence of oscillations. At the same
time we will fulfill the comparative analysis of the obtained
results at $CP$ violation in the above two approaches and also we
will fulfill comparison these results with the available
experimental data.

\section{$K^o_1, K^o_2$ meson vacuum oscillations at indirect
violation of $CP$ invariance with taking into account width
decays}

\par
The process of $K^o_1, K^o_2$ meson vacuum oscillations at
indirect violation of $CP$ invariance with taking into account
width decays in detail was considered in work \cite{13}. Therefore
we are considering main elements of these oscillations.

\par
It is clear that we have to take into account $CP$ phase $\delta$.
We can do it by using the parametrization of Kobayashi-Maskawa
matrix \cite{15} proposed by L. Maiani \cite{16}. The expressions
for $U, U^{-1}$ will then have the following form:
$$
U = \left(\begin{array}{cc} cos\beta & -sin\beta e^{-i \delta}\\
sin\beta e^{i \delta}& cos\beta \end{array} \right) \quad U^{-1} =
\left(\begin{array}{cc} cos\beta & sin\beta e^{-i \delta}\\
-sin\beta e^{i \delta} & cos\beta \end{array} \right). \eqno(1)
$$
Then at $CP$ violation $K^o_1, K^o_2$ mesons have to transform
into superposition states of $K_S$ and $K_L$ mesons
$$
\begin{array}{cc} K_S = {cos \beta K^o_1 - sin \beta
K^o_2} e^{-i \delta}, \\ K_L = { sin \beta e^{i \delta}K^o_1 + cos
\beta K^o_2},\end{array}
  \eqno(2)
$$
and at inverse transformation we get:
$$
\begin{array}{cc} K^o_1 = {cos \beta K_S + sin \beta e^{-i \delta}
K_L},\\ K^o_2 = {- sin \beta e^{i \delta} K_S + cos \beta K_L}
.\end{array}
  \eqno(3)
$$
In work \cite{13} it was shown that
$$
m_2 - m_1 \simeq m_L - m_S. \eqno(4)
$$
\par
If we take into account that $K_S, K_L$ decay and have the decay
widths $\Gamma_S, \Gamma_L$, then $K_S, K_L$ mesons with masses
$m_S$ and $m_L$ evolve in dependence on time according to the
following formula:
$$
K_S(t) = e^{-i E_S t - \frac{\Gamma_St}{2}} K_S(0), \qquad K_L(t)
= e^{-i E_L t- \frac{\Gamma_L t}{2}} K_L(0) , \eqno(5)
$$
where
$$
E^2_{k} = (p^{2} + m^2_{k}), \quad k = S, L  .
$$
If these mesons are moving without interactions, then
$$
\begin{array}{c}
K^o_1(t) = cos \beta e^{-i E_S t - \frac{\Gamma_S t}{2}} K_S(0) +
sin \beta e^{-i\delta}
e^{-i E_L t - \frac{\Gamma_L t}{2}} K_L(0) , \\
K^o_2(t) = - sin \beta e^{i \delta} e^{-i E_S t - \frac{\Gamma_S
t}{2}} K_{S}(0) + cos \beta e^{-i E_L t - \frac{\Gamma_L t}{2}}
K_L(0) .
\end{array}
\eqno(6)
$$
\par
Then using expressions (6) and (3) the probability that meson
$K^o_1$ produced at moment $t = 0$ will be at moment $t \neq 0$ in
the state of $K^o_2$ meson is given the following expression
$$
P(K^o_2 \rightarrow K^o_1, t) =  {1\over 4} cos^2 \beta \sin^{2}
2\beta [e^{-\Gamma_S t} + e^{-\Gamma_L t} -2 e^{-\frac{(\Gamma_S +
\Gamma_L) t}{2}} cos ((E_L - E_S) t) ] \eqno(7),
$$
If suppose that $cos^2 \beta \simeq 1$ and $sin^2 \beta \simeq
\varepsilon$ then
$$
P(K^o_2 \rightarrow K^o_1, t) \simeq \varepsilon
\left[e^{-\Gamma_S t} + e^{-\Gamma_L t} - 2 e^{-\frac{(\Gamma_S +
\Gamma_L) t}{2}} cos ((E_L - E_S) t)\right ] , \eqno(8)
$$
and $P(K^o_2 \rightarrow K^o_1, t) =  P(K^o_1 \rightarrow K^o_2,
t)$.
\par
Then the probability that meson $K^o_1$ produced at moment $t = 0$
will be at moment $t \neq 0$ in the state of $K^o_1$ meson and
back are given by the following expressions:
$$
P(K^o_1 \rightarrow K^o_1)  = [cos^4 \beta e^{-\Gamma_S t} + sin^4
\beta e^{-\Gamma_L t} +
$$
$$
 + 2 sin^2\beta cos^2 \beta e^{-\frac{(\Gamma_S +
\Gamma_L) t}{2}} cos ((E_L - E_S) t) ] , \eqno(9)
$$
further
$$
P(K^o_1 \rightarrow K^o_1) \simeq \left[e^{-\Gamma_S t} +
\epsilon^2 e^{-\Gamma_L t} + 2 \epsilon e^{-\frac{(\Gamma_S +
\Gamma_L) t}{2}} cos ((E_L - E_S) t) \right ] , \eqno(10)
$$
and probability $P(K^o_2 \rightarrow K^o_2)$ is
$$
P(K^o_2 \rightarrow K^o_2) = [sin^4 \beta e^{-\Gamma_S t} + cos^4
\beta e^{-\Gamma_L t} +
$$
$$
 + 2 sin^2\beta cos^2 \beta e^{-\frac{(\Gamma_S +
\Gamma_L) t}{2}} cos ((E_L - E_S) t) ], \eqno(11)
$$
further
$$
P(K^o_2 \rightarrow K^o_2) \simeq \left[\epsilon^2 e^{-\Gamma_S t}
+ e^{-\Gamma_L t} + 2 \epsilon e^{-\frac{(\Gamma_S + \Gamma_L)
t}{2}} cos ((E_L - E_S) t) \right ] .  \eqno(11')
$$
In the all above expressions we have to add factor $\frac{1}{2}$
since it arises from the primary $K^o, \bar K^o$ mesons ($K^o =
(K^o_1 + K^o_2)/\sqrt{2}$, $\bar K^o = (K^o_1 - K^o_2)/\sqrt{2}$
).
\par
When matrix transformation is unitary then $CP$ phase in the
expressions for transition probabilities is absent. In expression
(1) matrix $U$ is unitary, i. e., $U U^{-1} = 1$. In principle we
can use the non unitary matrix, i. e., to use matrix $U$ and for
back transformation to use matrix $U^{T}$ instead of $U^{-1}$
($detU = detU^{T} = 1$), then
$$
U = \left(\begin{array}{cc} cos\beta & -sin\beta e^{-i \delta}\\
sin\beta e^{i \delta}& cos\beta \end{array} \right) \quad U^{T} =
\left(\begin{array}{cc} cos\beta & sin\beta e^{i \delta}\\
-sin\beta e^{-i \delta} & cos\beta \end{array} \right). \eqno(12)
$$
Now instead of expr. (2) and (3) we get
$$
\begin{array}{cc} K_S = {cos \beta K^o_1 - sin \beta
K^o_2} e^{i \delta}, \\ K_L = { sin \beta e^{-i \delta}K^o_1 + cos
\beta K^o_2},\end{array}
  \eqno(13)
$$
$$
\begin{array}{cc} K^o_1 = {cos \beta K_S + sin \beta e^{-i \delta}
K_L},\\ K^o_2 = {- sin \beta e^{i \delta} K_S + cos \beta K_L}
.\end{array}
  \eqno(14)
$$
Now if mesons are moving without interactions, then
$$
\begin{array}{c}
K^o_1(t) = cos \beta e^{-i E_S t - \frac{\Gamma_S t}{2}} K_S(0) +
sin \beta e^{-i\delta}
e^{-i E_L t - \frac{\Gamma_L t}{2}} K_L(0) , \\
K^o_2(t) = - sin \beta e^{i \delta} e^{-i E_S t - \frac{\Gamma_S
t}{2}} K_{S}(0) + cos \beta e^{-i E_L t - \frac{\Gamma_L t}{2}}
K_L(0) .
\end{array}
\eqno(15)
$$
\par
Then using expressions (15) and (13) for the probability that
meson $K^o_1$ produced at moment $t = 0$ will be at moment $t \neq
0$ in the state of $K^o_2$ meson we get the following expression:
$$
\begin{array}{c} P(K^o_1 \rightarrow K^o_1) =\\
= \left[cos^4 \beta e^{-\Gamma_S t} + sin^4 \beta e^{-\Gamma_L t}
+  2 sin^2\beta cos^2 \beta e^{-\frac{(\Gamma_S + \Gamma_L) t}{2}}
cos ((E_L - E_S) t + 2 \delta) \right ]\end{array}, \eqno(16)
$$
or $sin^ \beta = \epsilon$ then
$$
P(K^o_1 \rightarrow K^o_1) \simeq \left[e^{-\Gamma_S t} +
\epsilon^2 e^{-\Gamma_L t} + 2 \epsilon e^{-\frac{(\Gamma_S +
\Gamma_L) t}{2}} cos ((E_L - E_S) t + 2 \delta) \right ] ,
\eqno(17)
$$
and the probability of $P(K^o_2 \rightarrow K^o_2)$ transition is
$$
\begin{array}{c}
P(K^o_2 \rightarrow K^o_2) = \\
 = \left[sin^4 \beta e^{-\Gamma_S t} + cos^4 \beta e^{-\Gamma_L t}
 + 2 sin^2\beta cos^2 \beta e^{-\frac{(\Gamma_S +
\Gamma_L) t}{2}} cos ((E_L - E_S) t + 2 \delta) \right
]\end{array} \eqno(18)
$$
or
$$
P(K^o_2 \rightarrow K^o_2) \simeq \left[\epsilon^2 e^{-\Gamma_S t}
+ e^{-\Gamma_L t} + 2 \epsilon e^{-\frac{(\Gamma_S + \Gamma_L)
t}{2}} cos ((E_L - E_S) t + 2 \delta) \right ] . \eqno(19)
$$
\par
Then the probability that meson $K^o_1$ produced at moment $t = 0$
will be at moment $t \neq 0$ in the state of $K^o_2$ meson is
given by the following expression:
$$
\begin{array}{ccc} P(K^o_2 \rightarrow K^o_1, t) = \\
= {1\over 4} \sin^{2} 2\beta [e^{-\Gamma_S t} + e^{-\Gamma_L t} -2
e^{-\frac{(\Gamma_S + \Gamma_L) t}{2}} cos ((E_L - E_S) t +
2\delta) ] \simeq \\ \simeq \varepsilon \left[e^{-\Gamma_S t} +
e^{-\Gamma_L t} - 2 e^{-\frac{(\Gamma_S + \Gamma_L) t}{2}} cos
((E_L - E_S) t + 2 \delta)\right ] ,
 \end{array} \eqno(20)
$$
and $ P(K^o_2 \rightarrow K^o_1, t) =  P(K^o_1 \rightarrow
K^o_2,t)$ (the above expression has taken into account that $cos^2
\beta \simeq 1$, $sin^2 \beta \simeq \epsilon)$.
\par
The length of oscillations in this case is
$$
R_{L S} \cong \frac{\gamma}{2 \Delta} \equiv \frac{2 \pi h c
\gamma}{2 \Delta} . \eqno(21)
$$
where $\Delta = m_L - m_S$ and $\gamma$ is usual relativistic
factor. The expressions (12) $\div$ (20) were obtained using the
standard technique of oscillations and then they are analogous to
the expression obtained in \cite{11, 12} at violation of
orthogonality of $K_S, K_L$ states.
\par
The graphics of transition probabilities $K^o_1 \to K^o_1$ (expr.
(10) -- $P(K^o, K^o_1 \to K^o_1, t) \simeq e^{-t} + (0.0023)^2
e^{-t/580} + 2 \cdot 0.0023 (cos(0.477t-0.752)) e^{-t
(581/1160)}$) and $K^o_2 \to K^o_1$ (expr. (8) -- $P(K^o, K^o_2
\to K^o_1, t) \simeq e^{-t} + (0.0023)^2 e^{-t/580}-2 \cdot 0.0023
(cos(0.477t-0.752)) e^{-t (581/1160)}$) in dependence on $t_S
=t/\tau_S$ ($\tau_S$ is $K_S$ life time) are given in Figure 1
(where $\varepsilon = 0.0023$ \cite{14}). And the summary graphic
of expressions (8) and (10) (line) normalized to the experimental
data from \cite{14} together with experimental data from \cite{14}
(open circles) is given in Figure 2 (for primary $K^o$ mesons).
From this figure we see that total transition probability to
$K^o_1$ obtained in the framework of oscillations theory are
placed very far from experimental data from \cite{14}. Then we can
come to the conclusion that at $CP$ violation in weak interactions
oscillations do not arise. In reality at drawing Figures 1, 2 it
was taken into account that there is phase $\delta= 44^o$ (i.e. we
used expressions (17) and (20)).

\begin{figure}[h!]
\begin{center}
\includegraphics[width=12cm]{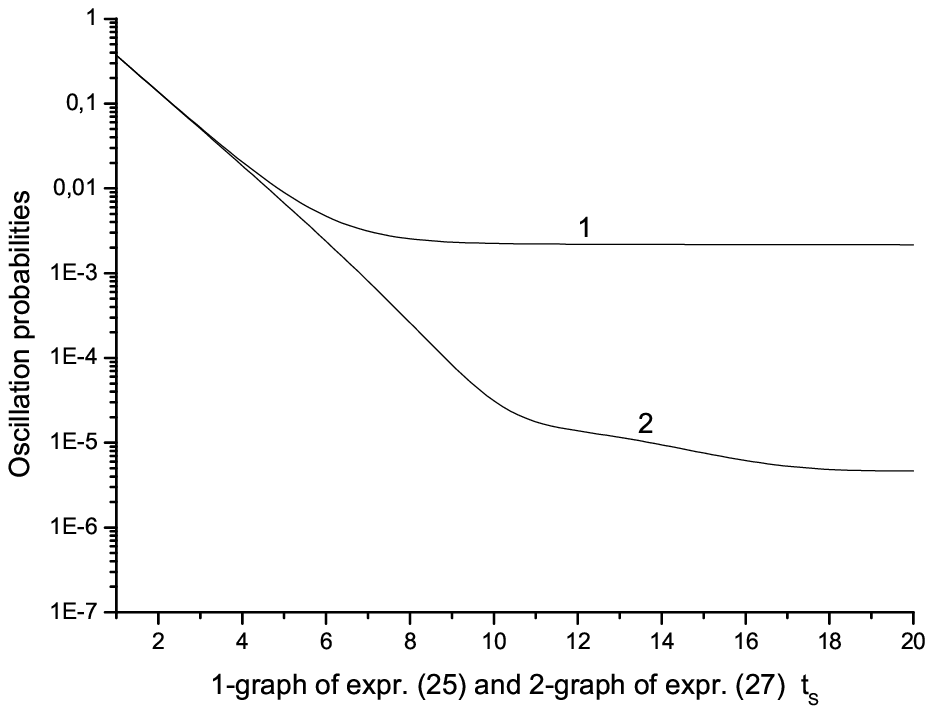}
\end{center}
\end{figure}

Figure 1. Line 1 is $K^o_2 \to K^o_1$ transition probability
(expression (8)) and line 2 is $K^o_1 \to K^o_1$ transition
probability (expression (10)) at presence of oscillations at $CP$
violation in weak interactions ($\varepsilon= 0.00223$) in
dependence on $t_S$ for $t_S = t/\tau_S = 1 \div 20$.\\

\newpage
\begin{figure}[h!]
\begin{center}
\includegraphics[width=12cm]{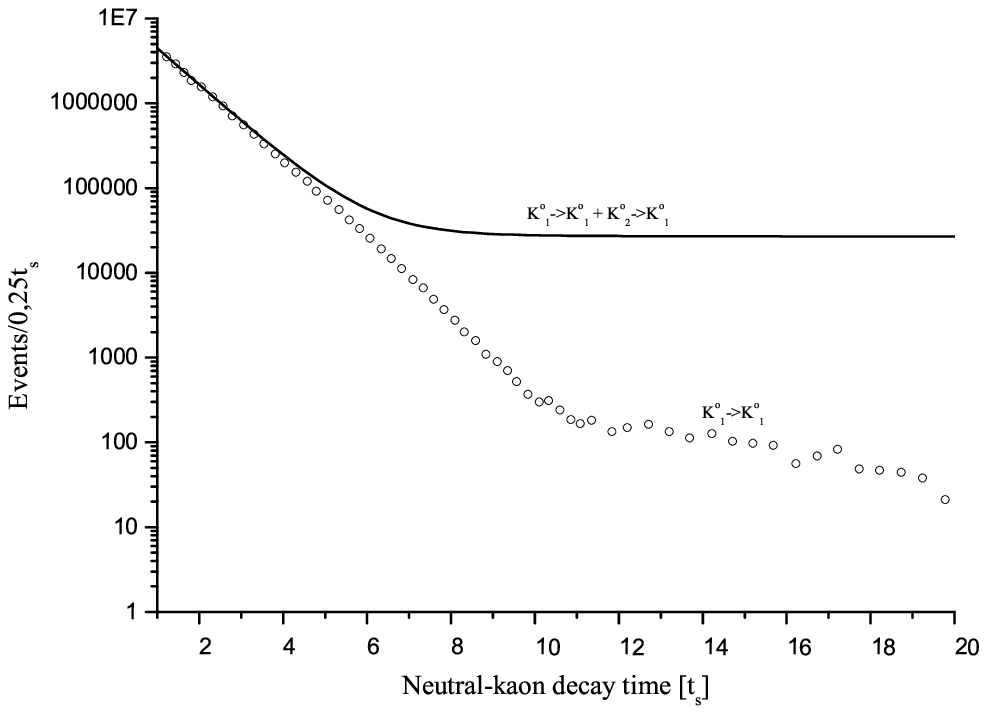}
\end{center}
\end{figure}

Figure 2. Summary transition probabilities ($K^o_1 \to K^o_1$) +
($K^o_2 \to K^o_1$) (line) when oscillations take place (exprs.
(8)+ (10)) normalized to experimental data from \cite{14} at
$t_S=1.22$ ($\varepsilon = 0.00223$) and
experimental data (open circles) from \cite{14} for $t_S = 1 \div 20$.\\

\par
Now we can consider the case when $\varepsilon' = \varepsilon^2 =
4.97 \cdot 10^{-6}$ then
$$
P(K^o, \bar K^o, K^o_1 \to K^o_1, t) = exp(-t) +
0.00000497(exp(-t)+
$$
$$
exp(-t/580) \pm 2(cos(0.477t-0.752))exp(-0.500862t)). \eqno(22)
$$
In Figure 3 there is presented line obtained by using the above
expression which normalized to the experimental data from
\cite{14} at $t_S=1.22$ and experimental data from \cite{14} for
$P(\bar K^o, K^o_1 \to K^o_1 , t\equiv t_S)$. We see that in this
case, the interference term which is present in the experimental
data, is absent. Then we can make the conclusion that oscillations
in this case also do not occur.

\newpage
\begin{figure}[h!]
\begin{center}
\includegraphics[width=9cm]{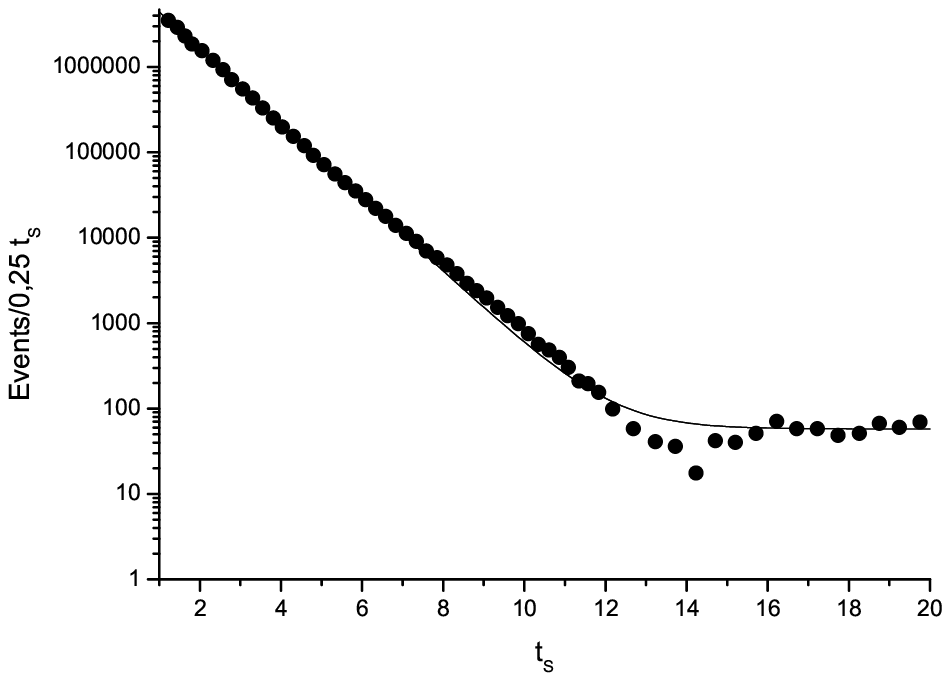}
\end{center}
\end{figure}

Figure 3. Summary transition probabilities ($K^o_1 \to K^o_1$) +
($K^o_2 \to K^o_1$) (line) when oscillations take place (exprs.
(8)+ (10)) normalized to experimental data from \cite{14} at
$t_S=1.22$ ($\varepsilon = 4.97 \cdot 10^{-6}$) and experimental
data (open circles) from \cite{14} for $t_S = 1 \div 20$.\\

\par
Now come to the consideration of the case when oscillations
between $K^o_1, K^o_2$ meson states do not arise at $CP$
violation.

\section{The case when at $CP$ violation oscillations between $K^o_1, K^o_2$ meson
states do not arise}

\par
Above we considered the case when at $CP$ violation there can
arise oscillations. Now we are considering the case when
superposition states arise but there are no oscillations. It
arises when there cannot be realized the condition for realization
of $K$ meson oscillations. Here an analogue with Cabibbo \cite{17}
mixing matrix takes place with one exclusion, namely, since masses
of $\pi$ and $K$ mesons differ very much then the interference
between these states  in contrast to $K_S, K_L$ meson states
cannot arise (by the way in full analogy with Cabibbo case we
could use below the old $K^o_1, K^o_2$ meson states instead of
using the new $K_S, K_L$ states).
\par
We know that the parameter of $CP$ violation is very small. Then
new states $K'_1 = cos \beta K_S + sin \beta K_L$ and $K'_2 =-sin
\beta K_S + cos \beta K_L$ are equivalent to $K^o_1, K^o_2$ states
($cos^2 \beta+ sin^2 \beta = 1$). Where $K_S, K_L$ states are
states which arise at small violation of $CP$ parity. They are not
orthogonal but normalized quantum mechanic function of states
($K_S(0)=1, K_L(0)=1$, $|K^o_1(0)|^2 + |K^o_2(0)|^2 = |K_S(0)|^2 +
|K_L(0)|^2$. Then
$$
\begin{array}{c}|K^o_1|^2 \equiv |K'_1|^2 = |cos \beta K_S + sin \beta K_L|^2,
\\ |K^o_2|^2 \equiv |K'_2|^2 =
|-sin \beta K_S + cos \beta K_L|^2, \\ |K'_1 K'_2|\simeq
0.\end{array} \eqno(23)
$$
As we see in this case instead of oscillations we get
interferences between $K_S$ and $K_L$ states. It is present
interest to rewrite the above expressions with taking into account
time dependence. Then taking into account that the standard
expressions for $K_S (t)$ and $K_L (t)$ have the following form:
$$
K_S (t) = exp^{(-i E_S t -\frac{1}{2} \Gamma_S t)} \quad K_L (t) =
exp^{(-i E_L-\frac{1}{2} \Gamma_S t)}, \eqno(24)
$$
and putting expressions (24) in (23) for a primary $K^o$ meson we
get
$$
|K'_1 (t)|^2 = cos^2 \beta exp^{(-\Gamma_S t)} + sin^2 \beta
exp^{(-\Gamma_L t)}
$$
$$
+ 2 sin \beta cos \beta exp^{(\frac{1}{2} (\Gamma_S+\Gamma_l) t)}
cos(E_L-E_S)t ,
$$
$$
|K'_2|^2 = sin^2 \beta exp^{(-\Gamma_S t)} + cos^2 \beta
exp^{(-\Gamma_L t)}
$$
$$
- 2 sin \beta cos \beta exp^{(\frac{1}{2} (\Gamma_S+\Gamma_l) t)}
cos(E_L-E_S)t , \eqno(25)
$$
$$
|K'_1 K'_2|\simeq 0.
$$
Since $K^o = \frac{1}{\sqrt{2}} (K^o_1 + K^o_2)$ then for the case
of a $K^o$ meson expressions (25) in normalized form get the
following form:
$$
|K'_1 (t)|^2 = \frac{1}{2}[cos^2 \beta exp^{(-\Gamma_S t)} + sin^2
\beta exp^{(-\Gamma_L t)}
$$
$$
+ 2 sin \beta cos \beta exp^{(\frac{1}{2} (\Gamma_S+\Gamma_l) t)}
cos(E_L-E_S)t ],
$$
$$
|K'_2|^2 = \frac{1}{2} [sin^2 \beta exp^{(-\Gamma_S t)} + cos^2
\beta exp^{(-\Gamma_L t)}
$$
$$
- 2 sin \beta cos \beta exp^{(\frac{1}{2} (\Gamma_S+\Gamma_l) t)}
cos(E_L-E_S)t] , \eqno(26)
$$
$$
|K'_1 K'_2|\simeq 0.
$$
For the case of a $\bar K^o$ meson we have
$$
\begin{array}{c}|K'_1|^2 = |cos \beta K_S - sin \beta K_L|^2, \\ |K'_2|^2 =
|sin \beta K_S + cos \beta K_L|^2, \\ |K'_1 K'_2|\simeq
0.\end{array} \eqno(27)
$$
Then using expression (24) for normalized case we get
$$
|K'_1 (t)|^2 = \frac{1}{2} [cos^2 \beta exp^{(-\Gamma_S t)} +
sin^2 \beta exp^{(-\Gamma_L t)}
$$
$$
- 2 sin \beta cos \beta exp^{(\frac{1}{2} (\Gamma_S+\Gamma_l) t)}
cos(E_L-E_S)t] ,
$$
$$
|K'_2|^2 = \frac{1}{2} [sin^2 \beta exp^{(-\Gamma_S t)} + cos^2
\beta exp^{(-\Gamma_L t)}
$$
$$
+ 2 sin \beta cos \beta exp^{(\frac{1}{2} (\Gamma_S+\Gamma_l) t)}
cos(E_L-E_S)t] , \eqno(28)
$$
$$
|K'_1 K'_2|\simeq 0.
$$
So, we have obtained the above expressions without the
renormalization of states by hand and without using non-unitary
matrix for transformation in contrast to work \cite{11}.
\par
It is present an interest the case when in expressions (23) will
be present a supplementary $CP$ phase. If this phase appears in
the unitary form as it is in \cite{15} in the form of \cite{16}
$$
U = \left(\begin{array}{cc} cos\beta & sin\beta e^{-i \delta}\\
-sin\beta e^{i \delta}& cos\beta \end{array} \right), \eqno(29)
$$
then in the case of $K^o$ meson instead of expressions (25) in the
case of $K^o$ meson we obtain:
$$
|K'_1 (t)|^2 = \frac{1}{2}[cos^2 \beta exp^{(-\Gamma_S t)} + sin^2
\beta exp^{(-\Gamma_L t)}
$$
$$
+ 2 sin \beta cos \beta exp^{(\frac{1}{2} (\Gamma_S+\Gamma_l) t)}
cos((E_L-E_S) + \delta) t ],
$$
$$
|K'_2|^2 = \frac{1}{2} [sin^2 \beta exp^{(-\Gamma_S t)} + cos^2
\beta exp^{(-\Gamma_L t)}
$$
$$
- 2 sin \beta cos \beta exp^{(\frac{1}{2} (\Gamma_S+\Gamma_l) t)}
cos((E_L-E_S)- \delta) t] , \eqno(30)
$$
$$
|K'_1 (t)|^2 \simeq \frac{1}{2}[exp^{(-\Gamma_S t)} +
\varepsilon^2 exp^{(-\Gamma_L t)} + 2 \varepsilon
exp^{(\frac{1}{2} (\Gamma_S+\Gamma_l) t)} cos((E_L-E_S)- \delta)
t] , \eqno(31)
$$
and in the case of $\bar K^o$ meson instead of expressions (26 )
we obtain:
$$
|K'_1 (t)|^2 = \frac{1}{2}[cos^2 \beta exp^{(-\Gamma_S t)} + sin^2
\beta exp^{(-\Gamma_L t)}
$$
$$
- 2 sin \beta cos \beta exp^{(\frac{1}{2} (\Gamma_S+\Gamma_l) t)}
cos((E_L-E_S) + \delta) t ] ,
$$
$$
|K'_2|^2 = \frac{1}{2} [sin^2 \beta exp^{(-\Gamma_S t)} + cos^2
\beta exp^{(-\Gamma_L t)}
$$
$$
+ 2 sin \beta cos \beta exp^{(\frac{1}{2} (\Gamma_S+\Gamma_l) t)}
cos((E_L-E_S)- \delta) t] , \eqno(32)
$$
$$
|K'_1 (t)|^2 \simeq \frac{1}{2}[exp^{(-\Gamma_S t)} +
\varepsilon^2 exp^{(-\Gamma_L t)} - 2 \varepsilon
exp^{(\frac{1}{2} (\Gamma_S+\Gamma_l) t)} cos((E_L-E_S)- \delta)
t] , \eqno(33)
$$
where using the existing experimental data \cite{14} we can write
that the value for $sin \beta$ is about $sin \beta = \varepsilon
\cong 2.23 \cdot 10^{-3}$.
\par
In Figure 4 are given graphic of functions (31)-- $P(K^o \to K_1,
t) \simeq e^{-t} + (0.00223)^2 e^{-t/580} + 2 \cdot 0.00223
(cos(0.477t-0.752)) e^{-t (581/1160)}$ normalized to the
experimental data from \cite{14} at $t_S=1.22$ together with
experimental data from \cite{14} for $t_S = 1 \div 20$ ($t_S =
t/\tau_S$, $\tau_S$ is $K_S$ meson life time).
\par
In Figure 5 are given graphic of functions (33) -- $P(\bar K^o \to
K_1, t) \simeq e^{-t} + (0.00223)^2 e^{-t/580} - 2 \cdot 0.00223
(cos(0.477t-0.752)) e^{-t (581/1160)}$ normalized to the
experimental data from \cite{14} at $t_S=1.22$ together with
experimental data from \cite{14} for $t_S = 1 \div 20$ ($t_S =
t/\tau_S$, $\tau_S$ is $K_S$ meson life time).
\par
We see that the curves from expressions (31) and (33) are in quite
satisfactory agreement with the experimental data obtained in
\cite{14} at $\varepsilon \cong 2.23 \cdot 10^{-3}$.

\newpage
\begin{figure}[h!]
\begin{center}
\includegraphics[width=10cm]{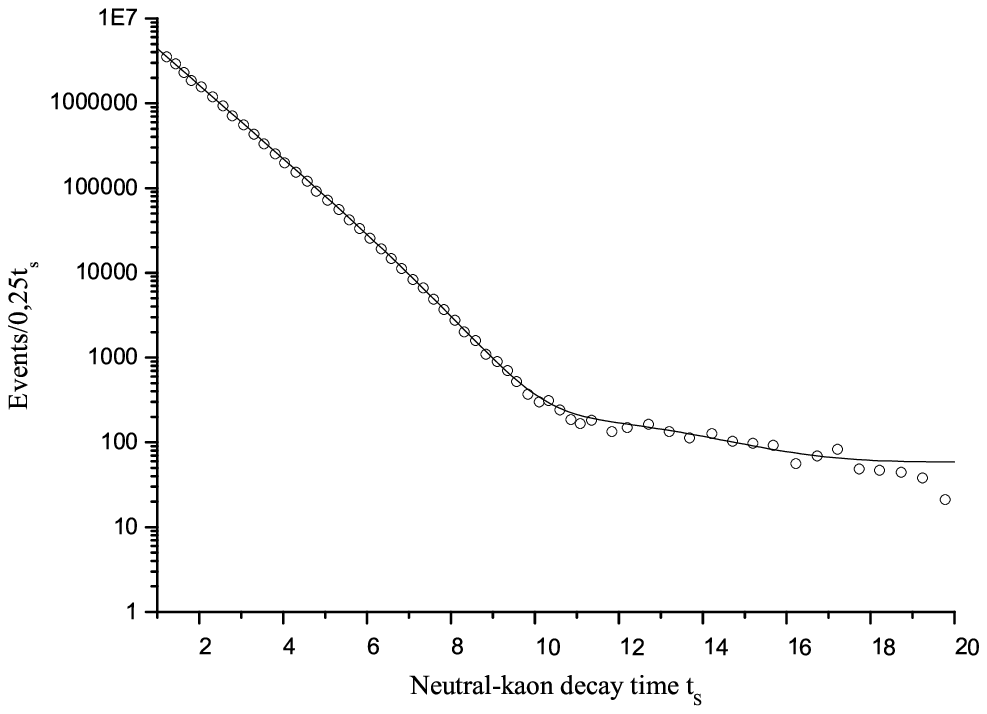}
\end{center}
\end{figure}

Figure 4. Transition probabilities of primary $K^o$ mesons into
$K_S$ ($P(K^o, K^o_1 \to K_S,t)$, expr. (31)) normalized to the
experimental data from \cite{14} at $t_S=1.22$ ($\varepsilon =
0.00223$) and
experimental data from \cite{14} for $t_S = 1 \div 20$.\\

\begin{figure}[h!]
\begin{center}
\includegraphics[width=10cm]{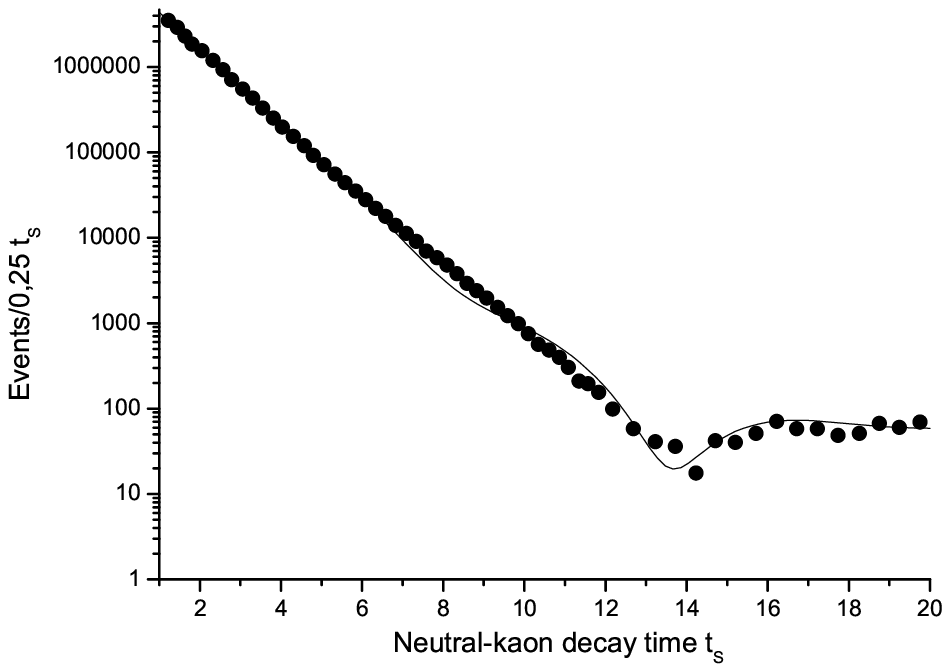}
\end{center}
\end{figure}

Figure 5. Transition probabilities of primary $K^o$ mesons into
$K_S$ ($P(\bar K^o, K^o_1 \to K_S,t)$, expr. (33)) normalized to
the experimental data from \cite{14} at $t_S=1.22$ ($\varepsilon =
0.00223$) and experimental data from \cite{14} for $t_S = 1 \div 20$.\\

\par
By the way, the signs of the additional $CP$ phase in our approach
are different for $K'_1$ and $K'_2$ mesons in contrast to
\cite{11} where was used non-unitary matrix transformation in the
case of $CP$ violation. Then there arises a question: what
mechanism works at $CP$ violation?  If it is possible to determine
this sign in experiment for a $K'_2$ meson then we can obtain the
answer to this question. If we use non-unitary matrix instead of
unitary matrix (29)
$$
U = \left(\begin{array}{cc} cos\beta & sin\beta e^{-i \delta}\\
-sin\beta e^{-i \delta}& cos\beta \end{array} \right), \eqno(34)
$$
then for $K^o$ and $\bar K^o$ transition probabilities we obtain
the same expressions that are in \cite{11}.
\par
So, as it has been was stressed above expressions for transition
probabilities (31), (33) are in good agreement with the
experimental data from \cite{14}. Then from expressions (31), (33)
and Figure 3, 4 we can come to conclusion that at $CP$ violation
in weak interactions the standard theory of oscillations is not
realized. There takes place only interference between two - $K_S,
K_L$ meson states.
\par
At $CP$ violation in weak interactions the mixing states of $K_S,
K_L$ mesons arise with very small angle mixing. These states are
not orthogonal states. I.e., there takes place an analogy with
Cabibbo matrix mixing \cite{17} at $\pi$, $K$ meson mixings with
one distinction: there arise interference between these states
since the masses of these states are very close. Then we can in
principle not introduce new $K_S, K_L$ states and use the old
$K^o_1, K^o_2$ meson states as it was done in the case of $\pi$,
$K$ mesons (or for $d$, $s$ quarks).

\section{Conclusion}

In this work have been considered two approaches for description
of $K^o, \bar K^o$ meson transitions into $K^o_1$ mesons at $CP$
violation in weak interactions. The first approach uses standard
theory of oscillations and the second approach supposes that
($K_S, K_L$) states which arise at $CP$ violation are normalized
but not but not orthogonal state functions, then there arise
interferences between these states but not oscillations between
them.
\par
In the case of  presence of oscillations the probability of $K^o,
\bar K^o$ meson transition into $K^o_1$ mesons is proportional to
$sin^2 \beta = \varepsilon = 2.23 \cdot 10^{-3}$ and at long
distances oscillations occur. In the second case there arises an
interference term between $K_S$ and $K_L$ meson states. This term
is proportional to $sin \beta = 2.23 \cdot 10^{-3}$ and it
disappears at big distances. And at big distances there is present
a term which is proportional to $sin^2 \beta = \varepsilon^2$. As
it was stressed above the available experimental data \cite{14}
are in good agreement with the second approach. So, we have come
to the conclusion that at $CP$ violation in weak interaction in
the system of $K^o$ mesons oscillations do not arise. There takes
place only interference between two - $K_S, K_L$ meson states.
\vspace{1cm}


\end{document}